\documentclass[iop,revtex4,twocolumn,nofootinbib,apj]{emulateapj}

\usepackage{amsfonts}
\usepackage{amsmath}
\usepackage{amssymb}
\usepackage{upgreek}
\usepackage{bm}
\usepackage{dcolumn}
\usepackage{epsfig}
\usepackage{graphicx}
\usepackage{placeins}
\usepackage{dsfont}
\usepackage{comment}

\newcommand{\thjn}{\theta_{\textrm{JN}}}
\newcommand{\thln}{\theta_{\textrm{LN}}}
\newcommand{\skyloc}{$\{\alpha, \delta\}$}
\newcommand{\skylocdist}{$\{\alpha, \delta, d_L\}$}
\newcommand{\skylocthjn}{$\{\alpha, \delta, \thjn\}$}
\newcommand{\skylocdistthjn}{$\{\alpha, \delta, d_L, \thjn\}$}

\newcommand{\stt}{\texttt{SpinTaylorT4}}
\newcommand{\imrp}{\texttt{IMRPhenomPv2}}

\newcommand{\nuaffil}{Center for Interdisciplinary Exploration and Research in Astrophysics (CIERA)
and
Department of Physics and Astronomy,
Northwestern University,
2145 Sheridan Road,
Evanston, IL 60208,
USA}

\begin{document}

\title{Astrophysical Prior Information and Gravitational-wave Parameter Estimation}

\author{Chris Pankow}
\author{Laura Sampson}
\author{Leah Perri}
\author{Eve Chase}
\author{Scott Coughlin}
\author{Michael Zevin}
\author{Vassiliki Kalogera}
\affiliation{\nuaffil}

\date{\today}

\begin{abstract}
The detection of electromagnetic counterparts to gravitational waves has great promise for the investigation of many scientific questions. While it is well known that certain orientation parameters can reduce uncertainty in other related parameters, it was also hoped that the detection of an electromagnetic signal in conjunction with a gravitational wave could augment the measurement precision of the mass and spin from the gravitational signal itself. That is, knowledge of the sky location, inclination, and redshift of a binary could break degeneracies between these extrinsic, coordinate-dependent parameters and the physical parameters that are intrinsic to the binary. In this paper, we investigate this issue by assuming perfect knowledge of extrinsic parameters, and assessing the maximal impact of this knowledge on our ability to extract intrinsic parameters. We recover similar gains in extrinsic recovery to earlier work; however, we find only modest improvements in a few intrinsic parameters --- namely the primary component's spin. We thus conclude that, even in the best case, the use of additional information from electromagnetic observations does not improve the measurement of the intrinsic parameters significantly.
\end{abstract}


\maketitle

\section{Introduction}
The Advanced LIGO detectors completed their first observing run in early 2016 (O1)~\cite{TheLIGOScientific:2016pea}. Within observations from O1, two gravitational wave (GW) events were confidently detected and determined to originate from two binary black hole (BBH) mergers: GW150914, observed
on September 14, 2015~\cite{PhysRevLett.116.061102} and GW151226, observed on December 26, 2015 ~\cite{PhysRevLett.116.241103}. A third candidate event, LVT151012, is not significant enough to be considered a detection, but is much more likely to be astrophysical in origin than a noise artifact ~\cite{Abbott:2016nhf}. The physical properties of these merger events were determined using Markovian sampling techniques~\cite{PhysRevLett.116.241102}. Electromagnetic (EM) observing partners did not report any sources definitively associated with any of these events~\cite{Abbott:2016gcq}, although~\cite{Connaughton:2016umz} reports on a gamma-ray event possibly associated with GW150914. 

The observation of a GW source and its EM counterpart --- for example from a neutron star-black hole (NSBH) binary --- would represent the first multimessenger event observed outside the local galactic neighborhood. It will not only directly inform scientists about questions such as the nature of short, hard gamma-ray bursts, but should also aid directly in GW parameter estimation~\cite{Bloom:2009vx,Nissanke:2009kt,Arun:2008zn}. 

GW signals from compact binary coalescences (CBC) are characterized by a set of intrinsic parameters (mass and spin of the binary components), and extrinsic, coordinate-dependent parameters~\cite{Maggiore2008} such as the luminosity distance $d_L$, inclination of the binary plane (the normal vector denoted $L$) with respect to the line of sight (vectorally denoted $N$, the associated tilt denoted $\thln$), and sky location (encoded in the antenna patterns $F_+, F_{\times}$ along with the polarization angle $\psi$):

\begin{widetext}
\begin{align}
\label{EqnWaveform}
\widetilde{h}_+(f) &= \frac{1}{d_L}F_+(\alpha, \delta, \psi)\left(\frac{1+\cos^2(\thln)}{2}\right) A(f)\exp(\Phi(f)) \nonumber \\
\widetilde{h}_{\times}(f) &= \frac{1}{d_L}F_{\times}(\alpha, \delta, \psi)\cos(\thln) A(f)\exp(\Phi(f) + \imath\pi/2)
\end{align}
\end{widetext}

\noindent Largely, the dependence of these two sets of parameters is decoupled since the dynamics of the binary, themselves dependent on the intrinsic parameters, are encoded in the frequency domain amplitude $A(f)$ and phasing $\Phi(f)$ of the gravitational waveform, and the extrinsic parameters determine the relative amplitude and mixing of the two polarizations. However, when one or both of the BH spins in the system is misaligned relative to the orbital angular momentum vector, the entire system precesses about the total angular momentum vector, and the inclination evolves over time. In almost all systems where $\thln$ is evolving due to precession, the tilt with respect to the \emph{total} angular momentum, (the vector denoted $J$, tilt denoted $\thjn$) changes very little over the evolution of the binary~\cite{PhysRevD.49.6274,PhysRevD.89.124025}, so this is the parameter we keep fixed.

Given information about the orientation parameters gleaned from an EM counterpart, it is possible to reduce degeneracies between various extrinsic parameters, and in principle, could lead to improved estimation of the remaining physical parameters from the GW signal. It is this effect that we seek to quantify. To analyze the utility of an EM counterpart on the estimation of GW source parameters, we assume that perfect knowledge of some extrinsic parameters (sky location, inclination, and/or distance via redshift measurements) is provided by an EM detection. Since at least a few NSBH~\cite{0264-9381-27-17-173001} are expected to be detected as advanced interferometers reach their design stage, the most likely EM counterpart to be detected in coordination with a gravitational-wave event is a short GRB. We perform a full Bayesian parameter estimation study of a population of NSBH sources using the LALInference pipeline~\cite{Veitch:2014wba} to assess the impact of this knowledge on extraction of intrinsic parameters. In short, we find that there is only weak improvement in the measurement of primary spin components when fixing the source orientation parameters --- other improvements remain statistically insignificant.

The rest of this paper is organized as follows. In Sec.~\ref{SecBackground} we give a brief introduction to expected EM counterparts and the information we should be able to extract from them. In Sec.~\ref{SecAnalysis} we describe our analysis, and in Sec.~\ref{SecResults} we present our results, with discussion following in Sec.~\ref{SecConclude}. 

\section{Information obtained from EM counterparts}
\label{SecBackground}

In a compact binary coalescence (CBC), a close binary composed of two neutron stars (BNS), two black holes, or a neutron star and a black hole spirals inward due to the emission of GWs, and eventually merges into a single object. It is expected that binaries which contain at least one neutron star will lead to EM signals due to the disruption of the neutron star matter. These signals are expected from a variety of sources, at a variety of wavelengths. A relativistic jet may lead to a short gamma-ray burst followed by X-ray, optical, and radio afterglows~\cite{1989ApJ...336..360E,1538-4357-561-2-L171,Berger:2007jk}. These events last from on the order of a second, to hours or days depending on wavelength. Kilonovae and macronovae in the optical and near-infrared range may be triggered by rapid neutron capture in ejecta~\cite{lattimer1976tidal,li1998transient}, and would last hours to weeks. Stellar-mass BHBH binaries, in contrast, are not expected to generate EM counterparts except for in quite exotic environments (although the detection in~\cite{Connaughton:2016umz} has resulted in a flurry of new proposed mechanisms~\cite{Zhang:2016rli,Yamazaki:2016fyr,Morsony:2016upv,Fraschetti:2016bpm,Malafarina:2016rdm,Janiuk:2016qpe,Liebling:2016orx,2041-8205-819-2-L21,2041-8205-821-1-L18,Stone01012017}).

Given the multitude of possibilities for EM signals corresponding to a GW trigger, what sort of advantages can we hope to leverage from an EM detection in constraining gravitational-wave event parameters? The most obvious is a good estimate of the sky location, which we will have from any EM counterpart. The sky localization of the aLIGO detectors is currently hundreds of square degrees (GW150914, for instance, was initially localized to an area of hundreds of deg$^2$~\cite{PhysRevLett.116.241102,2041-8205-826-1-L13}), whereas many EM telescopes will be able to localize the source to within a few square degrees. This improved localization can be useful in at least two different ways. In~\cite{Dalal:2006qt}, the authors explore how LIGO's search efficiency is improved when sky location is known due to the detection of a GRB --- a so-called triggered search. Additionally, in \cite{Holz:2002cn,Arun:2008zn}, it is shown that sky location information can greatly improve our ability to measure the luminosity distance to sources with LISA-type instruments.

In addition to sky localization, it is possible that certain classes of EM counterparts will give us information about the inclination of the binary relative to Earth, which in \cite{Nissanke:2009kt,1538-4357-688-2-L61} was shown to improve estimates of luminosity distance. Any EM source that is emitted in a jet geometry will, within the uncertainty of the beaming angle, provide just this kind of information~\cite{PhysRevLett.111.181101,Nakar2007166} --- see also~\cite{0004-637X-825-1-52} for an example of the geometry associated with kilonovae emission. Finally, if an EM counterpart can be identified with a host galaxy, it can provide independent information about the luminosity distance to the source via redshift measurements~\cite{PhysRevD.82.102002,Mandel_Kelley_Ramirez-Ruiz_2012}.

This systematic study is the first of its kind, but it only samples a small --- yet representative --- portion of the events that would be expected from NSBH with fully advanced detector configurations. We single out the NSBH source category as it is well studied and, given the electromagnetic energy emitted, the prime source expected to produce EM/GW coincidences in the next few years.

\section{Analysis Framework}
\label{SecAnalysis}

All parameter estimation in this study was performed using \texttt{lalinference\_mcmc}, the Markov Chain Monte Carlo parameter estimation code that belongs to the \texttt{LALInference} \cite{PhysRevD.91.042003} software library developed by the LIGO Scientific Collaboration and the Virgo Collaboration.

The injections we use in this study correspond to NSBH systems only --- it is derived from the distribution examined in \cite{Littenberg:2015tpa}. Source orientation parameters, such as sky location or inclination are all chosen isotropically. Black hole component masses ($m_1$) are uniformly distributed between 3 -- 30 $M_{\odot}$, and neutron stars ($m_2$) are uniformly distributed between 1 -- 3 $M_{\odot}$. Component spin vectors are isotropic  but distributed uniformly in magnitude ($a=|S/m^2|$). The injections are also placed uniformly in Euclidean volume, but were further down selected in order to obtain a sample of sources which had a signal-to-noise ratio (SNR, denoted by $\rho$) of at least 5 in the second highest SNR of the three detectors. This produces a SNR distribution $\propto \rho^{-4}$ above 5. It should be noted that while the power spectral density used in the likelihood is representative of the design sensitivity LIGO instruments, the MCMC runs themselves are analyzed in noise-free data.

We inject and recover systems using two families of waveform approximants: the $\stt$ family, a time-domain, inspiral-only, post-Newtonian approximant~\cite{PhysRevD.80.084043}, and the $\imrp$ family, a frequency-domain phenomenological family describing the full inspiral, merger, and ringdown~\cite{PhysRevD.91.024043,PhysRevLett.113.151101}. The $\imrp$ family is effectively limited to single-spin dynamics, however, we expect that this limitation is irrelevant to this analysis given the neutron-star spin is not expected to have a large influence because of relatively large mass ratios. Regardless, we cross check these results by processing the same systems using the $\stt$ family which includes full spin effects up to 2.5 post-Newtonian order. The $\imrp$ family contains only the primary $l=|m|=2$ modes, while the $\stt$ family also includes the $l=2, m=0$ mode. However, this mode is often several orders of magnitude smaller in amplitude to the dominant modes.

We use a lower frequency bound of 20 Hz near the boundary of accessible bandwidth expected for the era, and also use this frequency as reference point for the orientation of the binary relative to the line of sight. We use analytic marginalization of the likelihood over phase and time to coalescence.

For each of 91 binary merger events, we run five MCMC simulations, each individual run holding different sets of parameters fixed. Each represent different potential information sets that can be gleaned from an EM observation. We do not consider the effects of a noise realization --- the likelihood is calculated using a power spectral density representative of design sensitivity advanced LIGO~\cite{lrr-2016-1}, but the time series data itself is noise free. As such, we do not consider or the uncertainty introduced by imperfect calibration of data, but see \cite{PhysRevD.93.062002} for a discussion of how astrophysical prior information can be used to measure and constrain calibration uncertainties. 

\begin{enumerate}
\item none fixed --- Baseline comparison case, no information provided by electromagnetic observations
\item \skyloc~fixed --- Electromagnetic observatories have provided a location and likely an error region.
\item \skylocdist~fixed --- Observatories have provided a sky region and an estimate of distance, likely from redshift measurements. It is likely that EM observations will provide distance estimates with similar uncertainty to GW derived estimates. We ignore this and opt for the best case scenario where distance is known exactly.
\item \skylocthjn~fixed --- The event has been localized, and confirmed to be a collimated source (e.g. GRB). The opening angle of the jet is expected to be within 10-15${}^{\circ}$, so the approximation of exact knowledge is justified given the small difference in GW amplitudes across the allowable values.
\item \skylocdistthjn~fixed  --- Likely a GRB observed by a high energy event satellite and a measurable redshift.
\end{enumerate}

The goal of these simulations with different combinations of fixed parameters is to compare how precisely the various parameters of the binary can be measured given each set of information. The metric we use to gauge relative improvement in precision, for a given combination, is the area of the 90\% credible region. We examine the cumulative distribution of these areas over the same event population, and test whether the distributions deviate from our reference distribution (e.g. the one with no parameters fixed). Two tests are employed: we evaluate p-values for the two sample Kolmogorov--Smirnov (KS) test and also examine if the fixed parameter distribution lies within the bounds of the Dvoretzky-–Kiefer-–Wolfowitz (DKW) limit~\cite{dvoretzky1956} at the 95\% level from the empirical cumulative distribution function (CDF). These two measures are meant to test whether the statistical deviations of a given fixed parameter distribution is distinguishable from the unfixed distribution. In the case that this is true, the level of deviation (e.g. the p-value derived from the KS test) is indicative of how much better the parameter is measured with respect to no prior information. We can then determine the minimal amount of EM counterpart knowledge required in order to see improvement (if there is any) in the recovery of intrinsic parameters, such as component masses ($m_1, m_2$), spin ($a_1, a_2$), and spin tilt ($\theta_1, \theta_2$).

Finally, we note that the $\imrp$ is not recommended for use with spin configurations where $S$, the BH spin vector, is of similar magnitude and anti-aligned with $L$, the orbital angular momentum. Thus, we disallow extremely anti-aligned spin configurations, effectively where the spin of the black hole is greater than 0.9 of maximum.

\section{Results}
\label{SecResults}

\begin{figure*}[htbp]
\includegraphics[width=0.95\textwidth,height=4.5in]{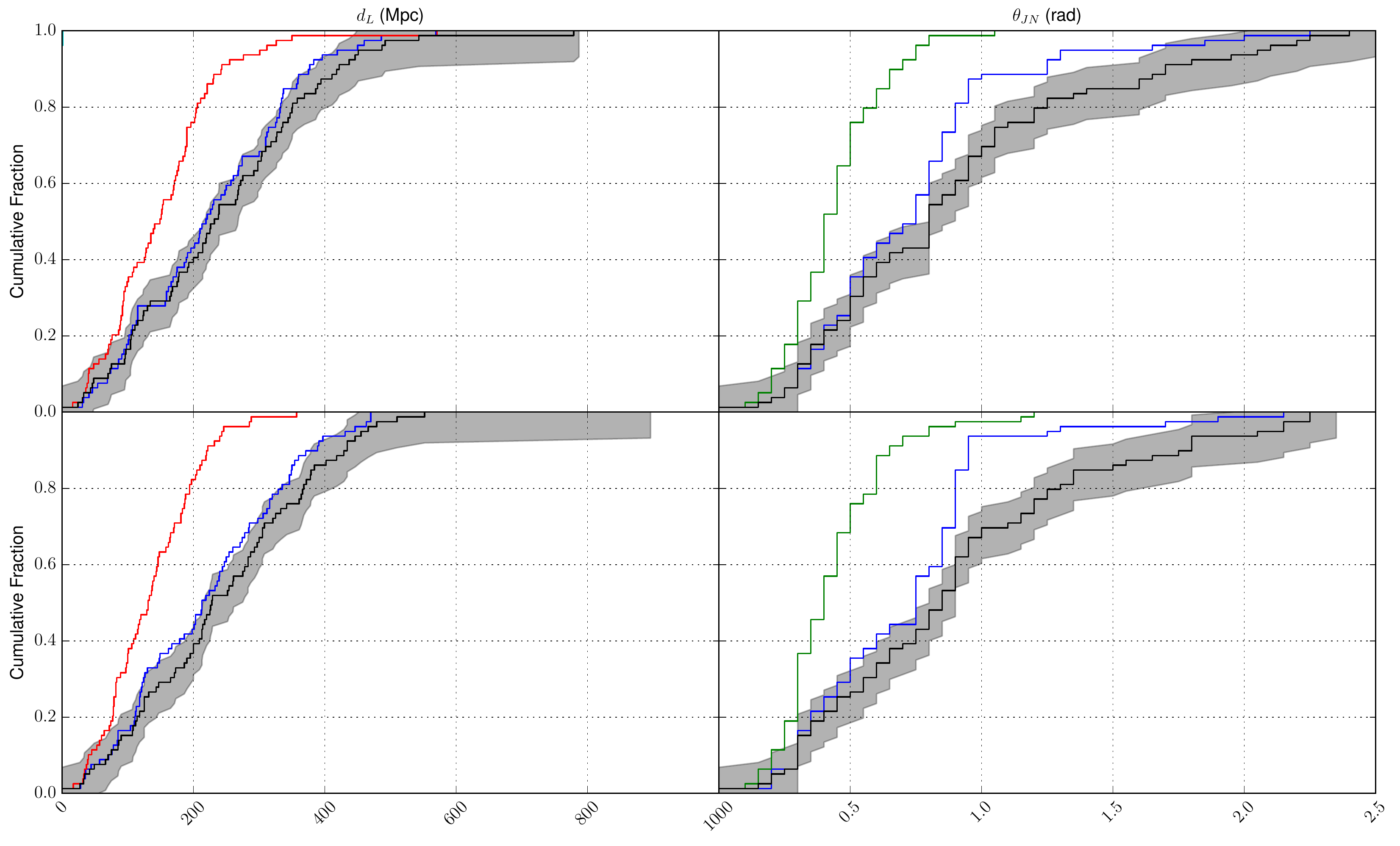}
\caption{\label{FigCDF1dextr} Cumulative fractions of the 90\% confidence regions for all source location parameters for the events examined in this study. The top panels correspond to the distributions using $\stt$ and the bottom corresponds to $\imrp$. The black curve is the reference unpinned distribution, with the grey error regions representing the DKW limit. The colored curves are the distributions for the various pinned distributions: sky location only is blue, sky location and distance is green, and sky location and inclination is red. Note that the plots corresponding to $\thjn$ have a coarse enough binning, to observe multiple events with the same quantized confidence interval value.}
\end{figure*}

We first consider the effect of the various parameter fixing on the other unpinned extrinsic parameters. Cumulative distributions for the extrinsic parameter ($\{\alpha, \delta\}, d_L, \thjn$) confidence intervals are shown in figure \ref{FigCDF1dextr}. Particularly, as the inclination and distance parameters appear as multiplicative factors in front of the intrinsic amplitudes in equation \ref{EqnWaveform}, they exhibit a very strong degeneracy. GW emission is beamed more strongly along the orbital angular momentum axis, so adjusting the overall amplitude by allowing $d_L$ to vary can be compensated by changing the viewing angle $\thln$ (and, by association $\thjn$). Stated simply, face on/off binaries ``appear'' closer than edge on binaries. As such, either distribution is markedly improved with the fixing of the other, sometimes even reducing the distance interval size by up to 50\% for binaries within the inspiral range~\cite{LIGO:2012aa} of the design sensitivity aLIGO era tested here. The relative improvement for $\thjn$ is of a similar order, with slightly more modest gains of $\sim 25-50$\% for larger absolute interval size when only the sky location is pinned. The distribution of $\thjn$ intervals is also improved by fixing the sky location, since the antenna factors also modify the overall amplitude of each polarization differently, and hence have strong covariance with the inclination factors.

\begin{figure*}[htbp]
\includegraphics[width=0.95\textwidth,height=4.5in]{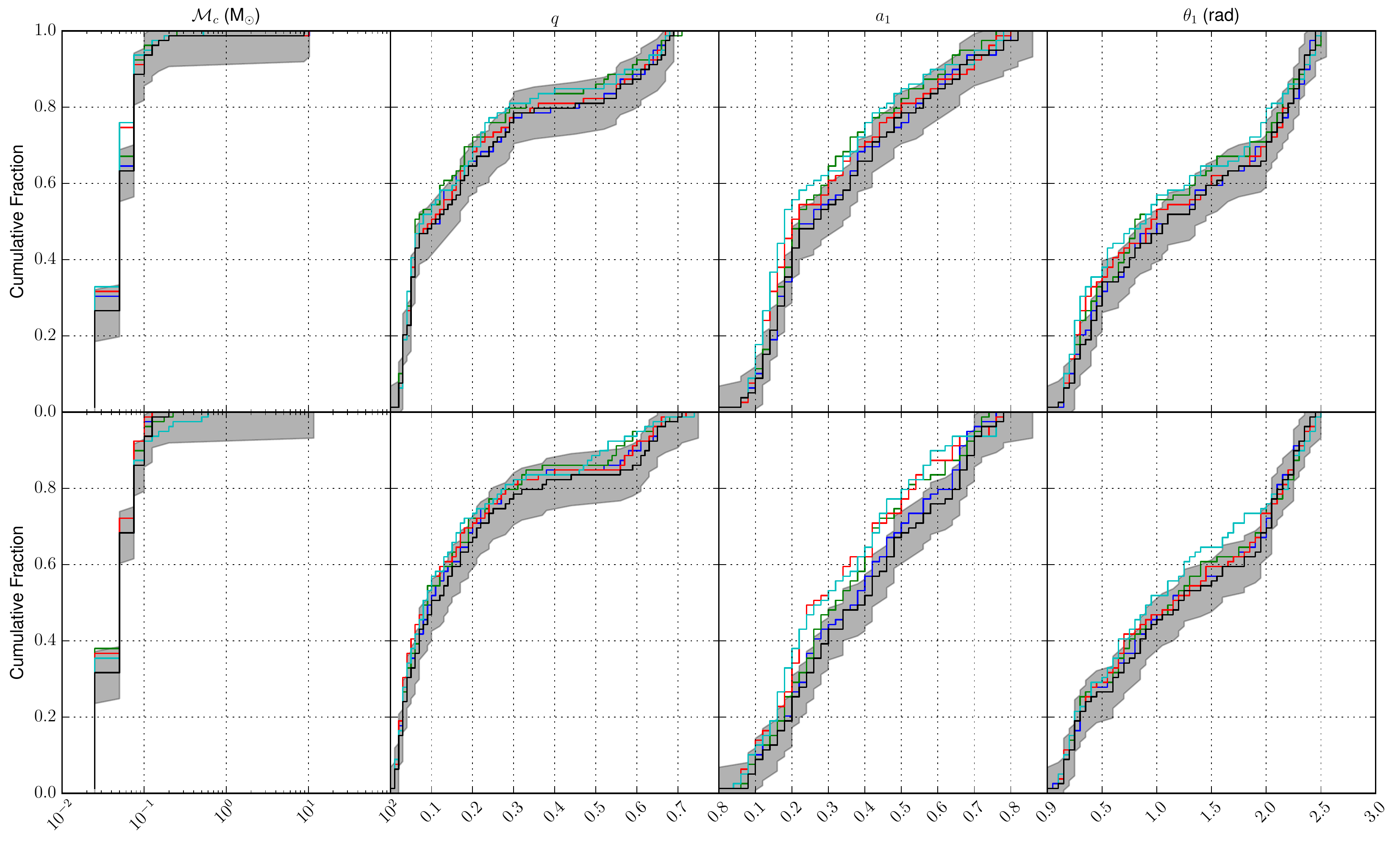}
\caption{\label{FigCDF1d} Cumulative fractions of the 90\% confidence regions for all source parameters for the events examined in this study. The positions and colors are exactly as in \ref{FigCDF1dextr}, with the addition of pinning all extrinsic parameters in cyan. The chirp mass distribution is log-scaled because its range of variation is otherwise indistinguishable.}
\end{figure*}

We examine the effect of the fixed parameter sets on the masses, here parameterized by chirp mass $\mathcal{M}_c = (m_1 m_2)^{3/5}/(m_1 + m_2)^{1/5}$ and mass ratio $q=m_2/m_1$), and also on the primary spin, here described by the dimensionless magnitude $a_1 = |S/m_1^2|$ and tilt relative to the orbital angular momentum $\cos\theta_1 = |\hat{L}\cdot\hat{S}_1|$. The azimuthal angle $\phi_1$ does not dramatically affect the precision with which we measure the other parameters). A comparison of the CDFs for single intrinsic parameter confidence intervals is presented in Figure \ref{FigCDF1d}. In almost all cases, the pinned distributions are not distinguishable from the unpinned distribution, either by the deviation within the DKW limit or from the p-value obtained from a KS test. The KS p-values are consistent with the observed deviations of the pinned distributions within the DKW error limits.

The most noticeable excursion occurs for the primary spin magnitude ($a_1$) with fixed $d_L$; Table \ref{TblSpinMagPVal} displays the KS p-values obtained for the various sets. The distributions are marginally outside the error region of the $\stt$ family for fixed sky location and distance and for all three fixed. 

In the case of $\imrp$, the situation is slightly better, with only the sky location pinning curve (blue) falling within the DKW error band. This may indicate that the addition of merger and ringdown information may help in measurement precision, since smaller regions are produced in cases where \emph{either} $d_L$ or $\thjn$ are fixed. 

\begin{table}[htbp]
\begin{tabular}{c|cccc}
family & \skyloc & \skylocdist & \skylocthjn & \skylocdistthjn \\
\hline
\stt & 1 & 0.66 & 0.8 & 0.16 \\
\imrp & 1 & 0.4 & 0.15 & 0.3 \\
\end{tabular}
\caption{\label{TblSpinMagPVal} KS p-values, relative to unpinned distribution, obtained for the primary spin magnitude parameter with a given fixed parameter set.}
\end{table}

\begin{figure*}[htbp]
\includegraphics[width=0.95\textwidth,height=4.5in]{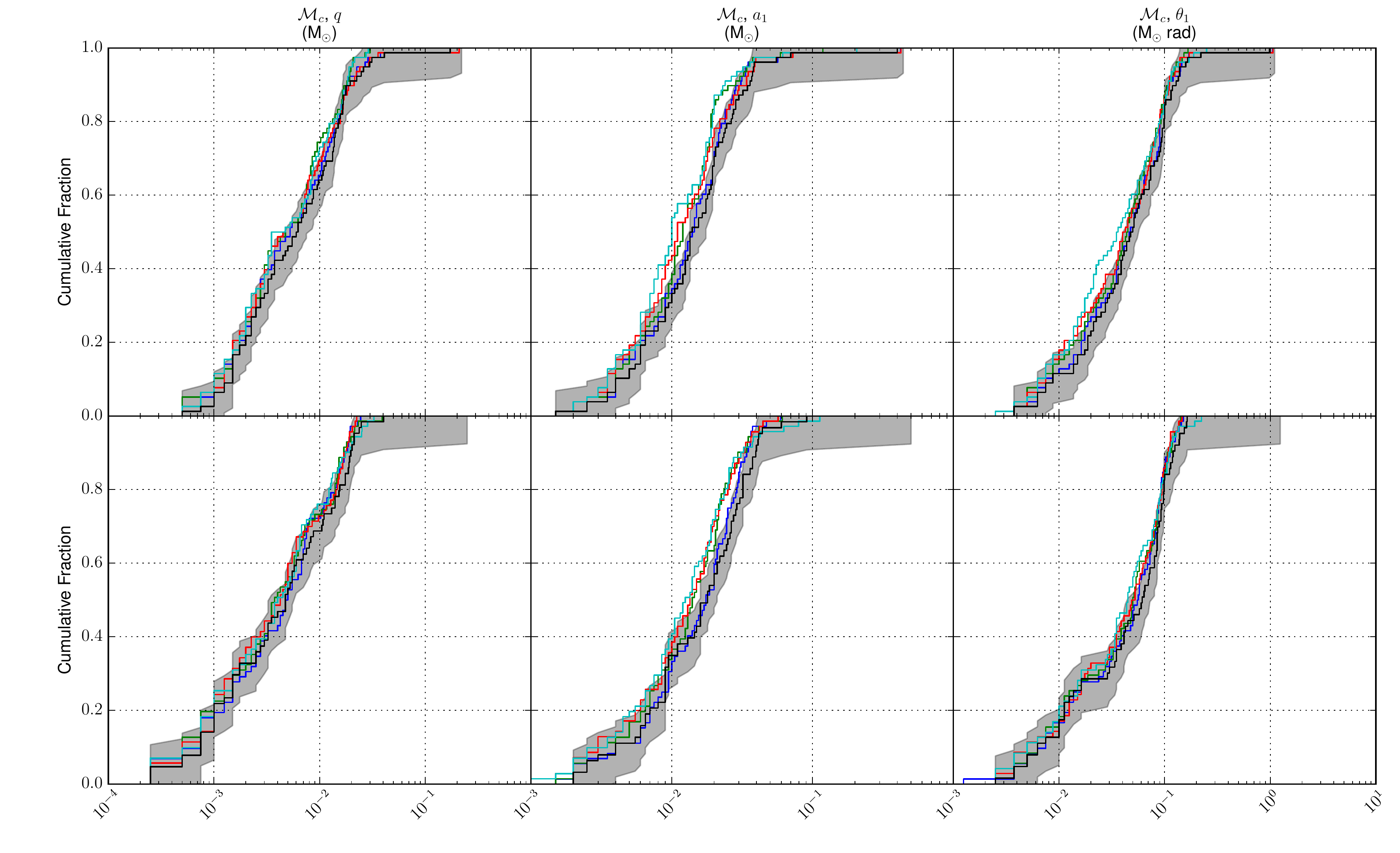}
\caption{\label{FigCDF2da}Cumulative distribution of the 90\% confidence regions for all two-dimensional source parameter combinations involving the chirp mass for the events examined in this study. The positions and colors are exactly as in Figure~\ref{FigCDF1d}. The chirp mass distribution is log-scaled because its range of variation is otherwise indistinguishable.}
\end{figure*}

\begin{figure*}
\includegraphics[width=0.95\textwidth,height=4.5in]{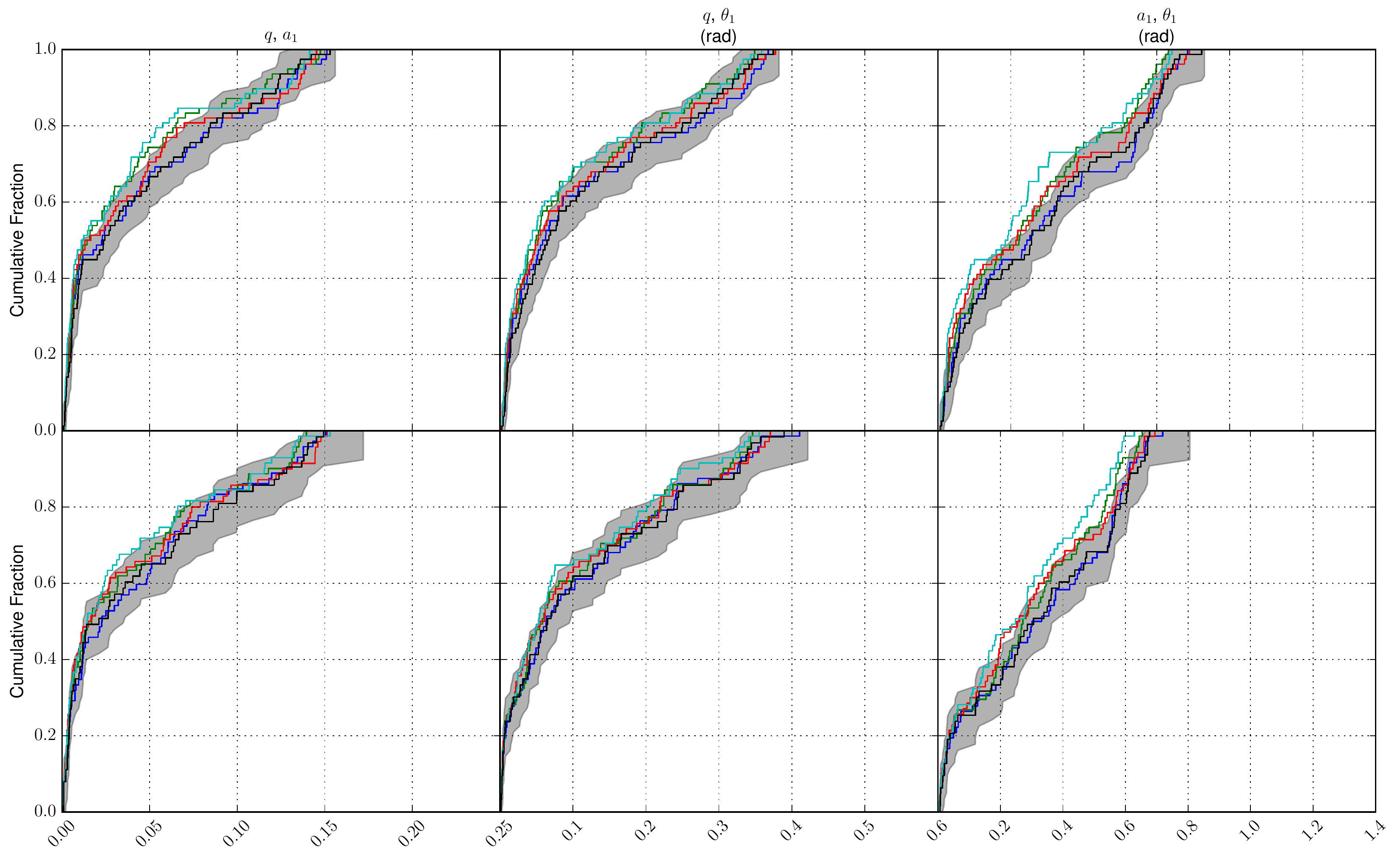}
\caption{\label{FigCDF2db}Cumulative distribution of the 90\% confidence regions for all remaining two-dimensional source parameter combinations for the events examined in this study. The positions and colors are exactly as in Figure~\ref{FigCDF1d}.}
\end{figure*}

Since many of the parameters in this study have significant correlations, it may also be useful to examine the effect of the EM prior on the area of the two-dimensional confidence region for pairs of parameters that we know to be correlated, for example, $\mathcal{M}_c$ and $q$, or $a$ and $q$. An overall reduction in the confidence region size distribution would indicate that while the one dimensional parameter may be unaffected, the correlated set has an overall reduction in posterior area. A selection of the two-dimensional area cumulative distributions are shown in Figures \ref{FigCDF2da} and \ref{FigCDF2db}. Again, in almost all cases, the distributions are not distinguishable from the unpinned reference distribution. Even in the case of ($a_1$, $\theta_1$), the smallest KS p-values are only 0.27 and 0.15 for $\imrp$ and $\stt$, respectively. While the secondary spin ($a_2$, $\theta_2$) distributions do have comparably deviating KS p-values in both the 1D and 2D cases, the posterior measured is not appreciably different than the prior, so the parameter was never ``measurable'' to begin with and any deviations are more likely due to random fluctuations in the recovered posteriors.

\section{Discussion and Conclusions}
\label{SecConclude}

Examining the most extreme deviations of the posterior area distributions, we find that there is only modest improvement in our ability to measure the intrinsic parameters when using the posterior area as our metric. Neither one nor two dimensional confidence region size were significantly affected by fixing the various orientation and distance parameters. There may be some additional precision obtained in the primary spin (most notably the primary spin magnitude), however, this requires knowledge of both the sky location and either the distance or relative inclination of the system. This is likely due to a strong correlation between $\thjn$ and the depth of the waveform amplitude modulation observed from a precessing system which is, in turn, dependent on both the spin magnitude and tilt.

Qualitatively, $\imrp$ and $\stt$ exhibit similar cumulative distributions for the same event set. It is notable, but not definitive, that the improvements in region size occur at absolute region sizes. For instance, the primary tilt angle is measured better with $\stt$, but only when the region is already relatively well contained within less than a quarter of the prior area. Conversely, $\imrp$ performs better over a wider set of regions. The deviations, while marginal, do occur for slightly different combinations of fixed parameters between the two families which may indicate that information from the merger and ringdown have some effect on our ability to measure certain parameters (e.g. spin orientation and magnitude).

This result reinforces the generally known result that the intrinsic source physical parameters and the source orientation parameters are mostly decoupled. Even in the case of precessing binaries where the spins and inclination are correlated because of the oscillation of the binary plane, we find here that having information about the inclination of the binary does not translate into noticeably better estimation of the spin parameters. Moreover, this result represents the best possible scenario in regards to incorporating priors from external astronomical information --- it effectively introduces a prior which is a delta function. In practice, the prior information will have a finite width and many gains made here will likely be lost.

Finally, we do note that, while not explored in detail here, the overall run times of the codes are reduced when fixing parameters. The effective decrease in dimensionality will improve convergence time in the MCMC used for measurement of the posteriors. Empirically, \imrp~was reduced by a factor of two between the unpinned case and the case fixing only sky location. Other configurations converged only slightly faster than the sky location case. The \stt~family configurations were affected in a similar way, but we observed a little less than a factor of two. For most compact binary sources, we expect that this procedure will be beneficial. Thus, this could be useful in reducing run time, for example like those reported in~\cite{Farr:2015lna}.

Moving forward into the era of joint gravitational wave and electromagnetic observations, our study shows that when targeting the measurement of the physical parameters of the binary there is no material benefit to incorporating event-by-event based astrophysically motivated priors to parameter estimation programs.

The authors are supported by NSF grants PHY-1307020 and PHY-1607709. This research was supported in part through the computational resources and staff contributions provided for the Quest high performance computing facility at Northwestern University which is jointly supported by the Office of the Provost, the Office for Research, and Northwestern University Information Technology. Specifically, we acknowledge  computing resources at CIERA funded by NSF PHY-1126812. Additionally, the authors would like to thank Ben Farr, Tyson Littenberg, Christopher Berry, Matt Pitkin, and Ray Frey for insightful commentary on the manuscript.

\bibliography{master}
\end{document}